\documentclass[useAMS,usenatbib,onecolumn]{mn2e}
\usepackage{epsfig}
\usepackage{amssymb}
\usepackage{epsfig,times}
\def\mincir{\raise -2.truept\hbox{\rlap{\hbox{$\sim$}}\raise5.truept \hbox{$<$}\ }}
\def\mincireq{\hbox{\raise0.5ex\hbox{$<\lower1.06ex\hbox{$\kern-1.07em{\sim}$}$}}}
\def\magcir{\raise-2.truept\hbox{\rlap{\hbox{$\sim$}}\raise5.truept \hbox{$>$}\ }}
\def\gr{\kern 2pt\hbox{}^\circ{\kern -2pt K}} 

\def\_{\thinspace}

\def\apj{{\it ApJ \,}}

\def\aj{{\it Astron. J.} \,}

\def\ni{\noindent}

\def\bs{\bigskip}

\def\ea{\ et al. \,}

\def\be{\begin{equation}}
\def\ee{\end{equation}}

\title[High Energy Emission from NGC\,253]{High Energy Emission from the Starburst 
Galaxy NGC\,253}
\author[Y. Rephaeli,Y. Arieli, M. Persic]{Yoel Rephaeli$^{1,2}$\thanks{E-mail:
yoelr@wise.tau.ac.il}, Yinon Arieli$^{1}$ and Massimo Persic$^{3}$\\
$^{1}$School of Physics and Astronomy, Tel Aviv University, Tel Aviv, 69978,
Israel\\
$^{2}$Center for Astrophysics and Space Sciences, University of California,
San Diego, La Jolla, CA 92093-0424\\
$^{3}$INAF/Osservatorio Astronomico di Trieste and INFN-Trieste, via G.B.Tiepolo 
11, 34143 Trieste, Italy}

\begin{document}
\pagerange{\pageref{firstpage}--\pageref{lastpage}} \pubyear{2007}

\maketitle


\label{firstpage}

\begin{abstract}
Measurement sensitivity in the energetic $\gamma$-ray region 
has improved considerably, and is about to increase further in the 
near future, motivating a detailed calculation of high-energy (HE: 
$\geq$100\,MeV) and very-high-energy (VHE: $\geq$100\,GeV) 
$\gamma$-ray emission from the nearby starburst galaxy NGC\,253. 
Adopting the convection-diffusion model for energetic electron and 
proton propagation, and accounting for all the relevant hadronic and 
leptonic processes, we determine the steady-state energy distributions 
of these particles by a detailed numerical treatment. The electron 
distribution is directly normalized by the measured synchrotron radio 
emission from the central starburst region; a commonly expected 
theoretical relation is then used to normalize the proton spectrum 
in this region. Doing so fully specifies the electron spectrum 
throughout the galactic disk, and with an assumed spatial profile of 
the magnetic field, the predicted radio emission from the full disk 
matches well the observed spectrum, confirming the validity of our 
treatment. The resulting radiative yields of both particles are 
calculated; the integrated HE and VHE fluxes from the entire disk 
are predicted to be $f(\geq 100\, {\rm MeV}) \simeq (1.8^{+1.5}_{-0.8})
\times 10^{-8}$ cm$^{-2}$s$^{-1}$, and $f(\geq 100\, {\rm GeV}) 
\simeq (3.6^{+3.4}_{-1.7})\times 10^{-12}$ cm$^{-2}$ s$^{-1}$, 
with a central magnetic field value $B_{0} \simeq 190 \pm 10$ $\mu$G. 
We discuss the feasibility of measuring emission at these levels with 
the space-borne {\it Fermi} and ground-based Cherenkov telescopes.
\end{abstract}

\begin{keywords}
galaxies:cosmic rays -- galaxies:gamma-ray -- galaxies:spiral -- 
galaxies:star formation
\end{keywords}

\section{Introduction}

High star formation (SF) and supernova (SN) rates in starburst (SB) 
galaxies (SBGs) enhance the energy density of energetic nonthermal 
particles - mostly electrons and protons - which are accelerated by 
SN shocks. Coulomb, synchrotron and Compton energy losses by the 
electrons, and the decay of pions following their production in 
energetic proton interactions with the ambient gas, result in emission 
over the full electromagnetic spectrum, from radio to high-energy (HE: 
$\geq$100\,MeV) $\gamma$-rays. The relatively high level of emission 
in SBGs (as compared with emission form `normal' galaxies) suggests 
nearby SBGs as viable candidates for detection by $\gamma$-ray 
facilities, such as the space-borne {\it Fermi} telescope, and the 
imaging air Cherenkov telescopes (IACTs) H.E.S.S., MAGIC, and 
VERITAS (e.g., De Angelis et al. 2008), all of which have already 
detected emission from a sample of AGN. When $\gamma$-ray emission 
is detected from SBGs, important additional insight will be gained on 
the origin and propagation mode of energetic electrons and protons in 
these galaxies. However, only weak level of emission is expected, making 
nearby SBGs the obvious choice of our study and potential targets for 
observations. A realistic estimate of the expected $\gamma$-ray 
emission requires a detailed account of all relevant energy loss 
processes of energetic electrons and protons as they move out from the 
central SB source region into the outer galactic disk. 

Measurements of very-high-energy (VHE: $\geq$100\,GeV) emission 
from NGC\,253 have so far resulted only in an upper limit obtained from 
observations with the H.E.S.S (Aharonian \ea 2005) Cherenkov telescopes. 
This limit is close to previous estimates of the HE $\gamma$-ray emission 
from this galaxy (Goldshmidt \& Rephaeli 1995, Romero \& Torres 2003, 
Domingo-Santamaria \& Torres 2005). The latter authors calculated HE 
$\gamma$-ray emission from the SB region of NGC\,253 based solely on 
spectral description of the electron and proton distributions. Thus, 
on both observational and theoretical grounds, and the additional 
motivation provided by improved observational capabilities afforded by 
the expected upgrade of Cherenkov telescopes, there is considerable 
interest in a more realistic treatment of the full nonthermal (NT) 
emission from NGC\,253. Moreover, recent detailed radio measurements 
of this galaxy (Heesen \ea 2008) clearly show the spatial distribution 
of the emission and its spectral index in the central region, disk, 
halo, providing an improved observational basis upon which the spectral 
and spatial distributions of the radio emitting electrons can be 
determined more precisely than was previously possible. These 
considerations spur our renewed interest in a substantially more 
detailed description of the electron and proton populations and their 
radiative yields.

Our main goal in the work reported here is obtaining an estimate of 
the HE flux of NGC\,253, more reliably than can be obtained 
by the traditional, very approximate approach, in which the electron 
energy spectrum is determined from the observed radio flux, and the 
proton spectrum is then deduced by using an overall scaling of the 
proton to electron ratio, which in turn is used to estimate the high 
energy emission from $\pi^0$ decay (following the production of the 
pions in energetic proton interactions with protons in the gas). 
However, in this simplified approach the electron spectrum needs to 
be extrapolated to energies much higher than the $\sim 1-10$ GeV 
range directly inferred from radio measurements; therefore, a 
realistic estimate of the VHE emission requires a more detailed 
and extensive treatment. For this purpose we use a numerical code 
which evolves an initial particle source spectrum in the SB 
acceleration region by following all the relevant leptonic and 
hadronic interactions as the particles diffuse and convect to 
the outer disk (and halo). In the first application of this code 
we calculated the $\gamma$-ray spectrum of the nearby SBG M\,82 
(Persic, Rephaeli, \& Arieli 2008: hereafter PRA). A comparison 
of the predicted spectrum with the predictions from approximate 
treatments (previous as well as our own) clearly demonstrates the 
need for a detailed numerical calculation. 

This paper presents only the most essential aspects of our treatment; 
to avoid unnecessary repetition we do not repeat here full details 
of the basic approach which can be found in PRA. In Section 2 we 
briefly summarize the main aspects of our treatment of the evolution 
of energetic electron and proton spectro-spatial distributions as the 
particles diffuse and convect out of their acceleration sources in the 
central SB region. Input conditions in NGC\,253 are specified in 
Section 3, together with the steady state spectra and VHE fluxes as 
calculated numerically using our code. We end with a brief discussion 
in Section 4.

\section{Steady-State Treatment}

High-mass stars form at a rate that is much higher in the central 
SB region than in the rest of the galactic disk. Because of this  
we may simplify the treatment by assuming that direct particle 
acceleration is limited to the central SB region, which we refer 
to as the source region. Acceleration in SN shocks by the first-order 
Fermi process yields a power-law distribution with index $q$$=$2 (e.g., 
Protheroe \& Clay 2004) in a very wide energy range, from a value close to 
the mean thermal energy of the gas particles (in non-relativistic shocks) 
to a very high value ($\geq$10$^{14}$ eV). The accelerated 
proton-to-electron (p/e) density ratio, $N_{\rm p}/N_{\rm e}$, in the 
source region can be calculated assuming charge neutrality (Bell 
1978, Schlickeiser 2002). This ratio reaches its maximum value, 
$(m_{\rm p}/m_{\rm e})^{(q-1)/2}$ (for $q$$>$1), over most of 
the range of particle energies; $m_{\rm e}$ and $m_{\rm p}$ are 
the electron and proton masses. For the full dependence of this ratio 
on particle energy, and more discussion on this and other relevant 
physical processes, see PRA and references therein. 

The theoretically predicted value of the density ratio is valid in the 
source region, where energy equipartition is more likely to be attained 
since the relevant processes couple particles and fields more effectively 
than over the whole galactic disk. We thus infer $N_{\rm p}$ in the source 
region from $N_{\rm e}$, which in turn is deduced from radio measurements 
in the central SB region. Specifically, by adopting the theoretically 
expected expression $N_{\rm p}/N_{\rm e}=(m_{\rm p}/m_{\rm e})^{(q-1)/2}$ 
in the SB source region, we compute $N_{\rm p}$ from the electron density, 
which is itself determined from a comparison of the predicted radio spectrum 
with the observed spectrum. The fit provides both the normalization of the 
electron spectrum and the {\it actual} value of $q$, which is found to be 
larger than $2$ even in the central SB region. In this procedure the 
electron population is composed of both primary and secondary electrons, 
with the latter self-consistently determined by accounting for the pion 
yield of energetic protons with protons in the gas. A similar procedure 
was implemented in our similar numerical treatment in PRA. 
\footnote{Note that in their description of our similar analysis to estimate 
the VHE emission from M82, de Cea del Pozo, Torres, \& Rodriguez-Marrero (2009) 
state that we used a much higher value for the latter ratio. Neither this nor 
their claim that we did not properly include secondary electrons are correct.} 

As the particles move out of the acceleration region their respective 
energy distributions (`spectra') will evolve differently due to their 
different energy loss processes. The electron spectrum is more easily 
measured than that of protons due to their more efficient radiative 
losses, with electron synchrotron radio emission typically the most 
precisely measured. The electron spectrum can be inferred in the region 
where the radio emission is observed. This quantity can be related to 
the source spectrum through a solution of the kinetic equation describing 
the propagation mode and energy losses by electrons as they move out 
from their SN shock regions. The proton spectrum can then be determined 
based on the predicted $N_{\rm p}/N_{\rm e}$ ratio at their common source 
region.

Strictly speaking, we assume that steady state is attained and proceed 
to solve the kinetic equation for $N_{i}(\gamma, R,z)$, where $i=e, p$; 
$\gamma$ is the Lorentz factor, and $R$ and $z$ are the 2D spatial 
radius and the coordinate perpendicular to the galactic plane, 
respectively. Calculations of the particle steady-state spectra 
necessitate inclusion of all the important energy loss mechanisms and 
modes of propagation; to do so we have employed the numerical code of 
Arieli \& Rephaeli (in preparation), which is based on a modified 
version of the GALPROP code (Moskalenko \& Strong 1998, Moskalenko 
\ea 2003). The code solves the exact Fokker-Planck transport 
diffusion-convection equation (e.g., Lerche \& Schlickeiser 
1982) in 3D with given source distribution and boundary conditions 
for electrons and protons. Evolution of particle energy and spatial 
distribution functions is based on diffusion, and convection in a 
galactic wind. 

At high energies the dominant energy losses of electrons are 
synchrotron emission and Compton scattering by the FIR and optical 
radiation fields. These well known processes need no elaboration; the 
level of the ensuing emission depends on the mean strength of the 
magnetic field, $B$, and the energy density of the radiation fields, 
which are specified below. At energies below few hundred MeV, electrons 
lose energy mostly by Coulomb interactions with gas particles, leading 
to ionization of neutral and charged ions, and electronic excitations 
in fully ionized gas. 

At low energies proton losses are dominated by Coulomb interactions 
with gas particles. Protons with kinetic energy above pion masses 
($\sim$140 MeV) lose energy mainly through interactions with ambient 
protons, yielding neutral ($\pi^{0}$) and charged ($\pi^{\pm}$) pions. 
Neutral pions decay into photons, while decays of $\pi^{\pm}$ result 
in relativistic e$^{\pm}$ and neutrinos. In our treatment the electron 
component always includes both primary and secondary electrons; the 
latter are produced in $\pi^{-}$ decay following pion production in 
{\it pp} interactions of accelerated protons with interstellar gas.

\begin{table}
\caption{Measured values of NGC\,253 parameters.} 
\label{table:cluster}
\begin{center}
\small
\begin{tabular}{llll } \hline\hline
Parameter & Value & Units & Reference \\\hline
Distance & 2.5 & Mpc & 2 \\
SB region radius & 200 & pc & 6,7 \\
SB region height & 150 & pc & 6,7 \\
Disk radius & 10 & kpc &  \\
$M_{H2}(R<1.1$ kpc) & $4.8\times 10^8$ & $M_{\odot}$ & 8 \\
$M_{H}(R<600$ pc) & $4.8 \times 10^8$ & $M_{\odot}$ & 2 \\
$M_{H}(R<10$ kpc) & $2.5 \times 10^9$ & $M_{\odot}$ & 1 \\
$M_{HII}(R<6.35$ kpc) & $2 \times 10^7$ & $M_{\odot}$ & 5 \\
Dust temperature & 50 (SB), 16 (disk) & K & 3 \\
Dust emissivity index & 1.5 & &  3 \\
\end{tabular}
\end{center}

\ni{References to listed data: (1) Boomsma et al. 2005; (2) Bradford et al. 2003; 
(3) Melo et al. 2002; (4) Sreekumar et al. 1994; (5) Strickland et al. 2002; 
(6) Timothy et al. 1996; (7) Ulvestad 2000; (8) Canzian et al 1988.}

\end{table}

To determine the steady state distributions of electrons and protons we 
need to specify the densities of neutral and ionized gas in the central 
SB region and throughout the disk, properties of the magnetic fields, 
and the energy density of the ambient radiation field (in addition to the 
CMB), particularly in the IR region. Mean observed values of all these 
quantities are selected from the literature, including - when available 
- some knowledge of their spatial variation across the disk. 
Parameter values are listed in Table 1; observational 
uncertainties are taken into account in our estimation of the integrated 
VHE flux. As discussed in PRA, we assume magnetic flux conservation in 
the ionized gas, and accordingly relate the mean strength of the field 
to the local plasma density, $n_{\rm e}$, using the scaling 
$B \propto n_{\rm e}^{2/3}$ (e.g., Rephaeli 1988). The power-law 
index is only somewhat lower ($1/2$) if instead energy equipartition is 
assumed and the magnetic energy density is scaled to the thermal gas 
energy density. We adopt the ionized gas density profile, $n_{e} \propto 
exp(-z/z_0)/(1 + (R/R_0)^2)$, with $R_0=1.5$ kpc, and $z_0 =0.5$ kpc 
(Strickland et al. 2002). 

To fully determine particle spectra and the magnetic field strength, we 
assume energy equipartition in the source region. As noted above, we believe 
that a minimum energy condition is more likely to lead to equipartition in 
the central SB region than in the rest of the galaxy. 
Due to the implicit dependences in the expression for the synchrotron flux, 
this condition is implemented iteratively to solve for $N_{\rm e}$, 
$N_{\rm p}$, and the field strength at the center, $B_0$. The particles 
are assumed to be convected out of their sources by the observed wind 
with a typical velocity of $\sim 500$ km $s^{-1}$ (Strickland et al. 
1997) in the source region. In accord with Galactic cosmic-ray MHD 
wind models, we assume that the convection velocity increases linearly 
with distance from the disk plane (e.g. Zirakashvili et al. 1996). A 
central value of $3 \times 10^{28}$ cm$^2$/s is adopted for the diffusion 
coefficient, and its energy dependence is approximated as a power-law with 
index of $0.5$. 

For brevity (and to avoid unnecessary repetition), other aspects of our 
approach are not presented here; these are fully specified in our previous 
similar treatment of VHE emission from M\,82 (PRA). In Table 1 we list the 
values of all the quantities needed to carry out the full calculation of 
the steady-state particle spectra and the predicted radiative yields.

\section{Particle and Radiation Spectra}

Radio emission from the disk of NGC\,253 was observed at several 
frequencies by Klein et al. (1983); additional medium resolution 
measurements of the SB region were made by Carilli (1996). Detailed, 
higher resolution mapping 
of emission from the disk and inner halo of NGC\,253 was recently carried 
out by Heesen \ea (2008) based on new VLA observations at 6.2 cm, 
Effelsberg observations at 3.6 cm, and previous VLA measurements at 20 
and 90 cm. We normalize the electron source spectrum based on the 
measured radio flux from the SB region (Carilli 1996), and determine 
some of our best-fit parameters by comparing our predicted radio emission 
from the galactic disk with the spectrum measured by Heesen \ea (2008).

Fitting the predicted synchrotron radio spectrum to the radio measurements 
provides the critical normalization of the steady state electron and - 
based on theoretical prediction - proton energy distributions. Inclusion 
of all the relevant processes - Coulomb, synchrotron and Compton - is 
{\it essential} for precise normalization. Measured radio fluxes from the 
SB (Klein et al. 1983) and disk (Heesen \ea 2008) regions are shown in 
Figure 1, together with our best fit spectra. The fact that the predicted 
radio spectrum is not a single power-law but curves at sufficiently low 
frequencies is due to the dominance of Coulomb losses over the other two 
mechanisms at low electron energies. For the specific combination of 
values of the field and gas density (which we have deduced from 
observations), the curvature is particularly pronounced (at around 
$\sim 3$ GHz) in the source region. This expected behavior is a 
diagnostically important feature confirming the self-consistency of 
our approach. The fit to the radio measurements from the full disk 
yields a mean radio index $\alpha = 0.7 \pm 0.05$.

\begin{figure}[h]
\label{fig:radio_spectrum}
\centering
\epsfig{file=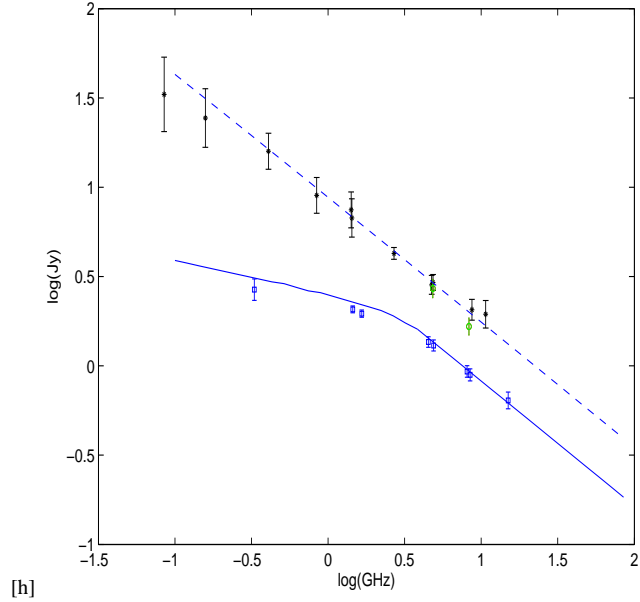,height=8cm,width=8cm,clip=}
\caption{Radio measurements of the SB and entire disk regions of NGC\,253. 
The solid line is our fit to the emission from the SB region; the dashed 
line is a fit to the emission from the entire disk. Data are from Klein 
et al. (1983, black dots), Carilli (1996, blue squares), and Heesen \ea 
(2008, green circles).}
\end{figure}

As noted above, energy equipartition was assumed and implemented 
iteratively to solve for $N_{\rm e}$, $N_{\rm p}$, and $B_0$. Doing 
so we obtain the central value of the magnetic field, 
$B_{0} \simeq 190 \pm 10$ $\mu$G. The corresponding steady-state 
electron and proton spectra in the SB region are shown in Fig.2. At low 
energies the spectra are appreciably flatter than at high energies. 
Stronger electron losses at $E$$>>$1 GeV result in a steeper spectrum 
than that of protons, with the electron spectrum characterized by 
$q \simeq 2.74$ at high energies, as compared $q \simeq 2.55$ for 
protons. We note that the validity of the usual equipartition relation 
between particle and field energy densities has been questioned (e.g., 
Beck \& Krause 2005), but in our case the revised relation proposed 
in the latter paper would hardly affect the deduced mean field 
value.

\begin{figure}[h]
\label{fig:ep_density}
\centering
\epsfig{file=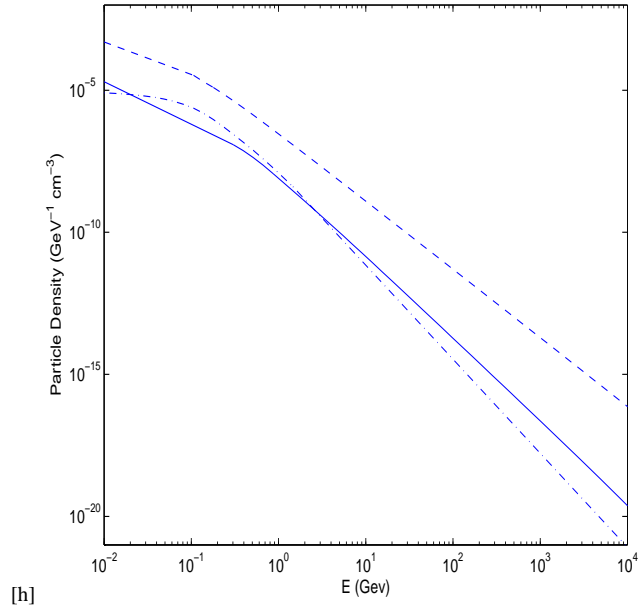,height=8cm,width=8cm,clip=}
\caption{Steady state primary proton (dashed line), primary electron 
(solid line), and secondary electron (dashed-dotted line) spectral 
density distributions in the central SB region NGC\,253.}
\end{figure}

Our main interest here is the estimation of the total emission at 
high $\gamma$-ray energies. 
In Fig. 3 we plot electron bremsstrahlung and Compton spectra, and 
$\gamma$-ray emission from the decay of $\pi^0$ formed in p-p 
collisions. As anticipated, the losses due to bremsstrahlung dominate at 
low energies, while emission from $\pi^0$ decay dominates at higher 
energies. (Synchrotron emission is negligible at high energies.) 

The integrated flux from the disk of NGC\,253 is shown in Fig. 4 by the 
grey region; the width of this region reflects uncertainties in the values 
of the main obesrvationallly-deduced parameters. The total flux above 
$100$ GeV is $f(\geq 100 \, {\rm GeV}) \simeq (3.6^{+3.4}_{-1.7})
\times 10^{-12}$ cm$^{-2}$ s$^{-1}$, for $B_{0} \simeq 190 \pm 
10$ $\mu$G, the deduced range of the radio index, $\alpha = 0.7 \pm 
0.05$, and the uncertainty in the ambient gas density. As expected, 
VHE emission comes mainly from the hadronic $\pi^0$ decay channel. 
Emission at these energies is mostly from the SB region, with the rest 
of the disk contributing $20\%-40\%$ of the total emission. 
The flux at $\epsilon \geq 50$ GeV is a factor of $\sim 3.6$ higher. 
Emission at much lower energies can be measured by the currently 
operational {\it Fermi} satellite; for example, we predict an 
integrated flux $f(\geq 100\, {\rm MeV}) \simeq (1.8^{+1.5}_{-0.8})
\times 10^{-8}$ cm$^{-2}$s$^{-1}$. This value matches the {\it Large 
Area Telescope} 5$\sigma$ sensitivity of this instrument for a 1-yr 
scanning-mode operation (e.g., Atwood \ea 2009). Of interest is also 
the related neutrino flux from the decay of charged pions $\pi^{\pm}$ 
(that produce $e^{\pm} + \nu_{e} + \bar{\nu}_{\rm e} + \nu_{\mu} + 
\bar{\nu}_{\mu}$): at $E \geq 100\,$GeV the neutrino flux is $\sim 0.3\, 
f(\geq 100\,{\rm GeV})$. 

Our exact numerical calculation yields flux levels that are lower than 
those estimated from a semi-quantitative calculation that does not 
include the full spatial dependences of the magnetic field 
and particle densities. Details of this approximate treatment are 
given in our previous paper (PRA) on VHE emission from M\,82. As noted 
there, the two treatments differ most in the description of the 
particle spatial profiles outside the source region. In the approximate 
treatment the impact of energy losses and propagation mode of the protons 
are not explicitly accounted for. This leads to unrealistically high 
relative contribution of the main disk to VHE emission. Even a slight 
steepening of the proton spectrum at high energies results in a 
significantly lower VHE emission.

\begin{figure}[h]
\label{fig:spectra_all}
\centering
\epsfig{file=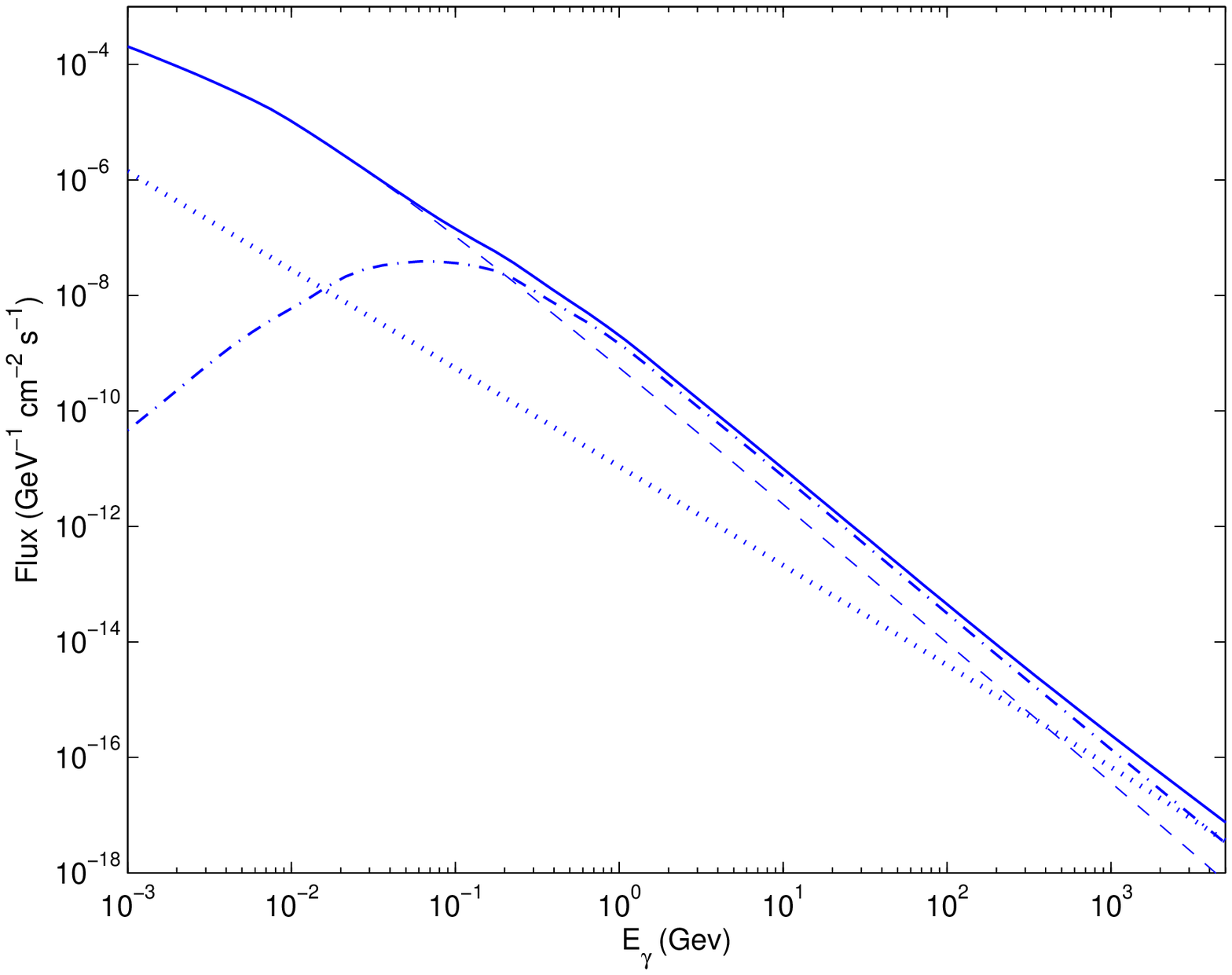,height=8cm,width=8cm,clip=}
\caption{Spectra of high-energy emission processes in the disk region 
of NGC\,253. Radiative yields are from electron Compton scattering off 
the FIR radiation field (dotted line), electron bremsstrahlung off 
ambient protons (dashed line), $\pi^0$ decay (dashed-dotted line), and 
their sum (solid line).}
\end{figure}

\begin{figure}[h]
\label{fig:integrated_flux}
\centering
\epsfig{file=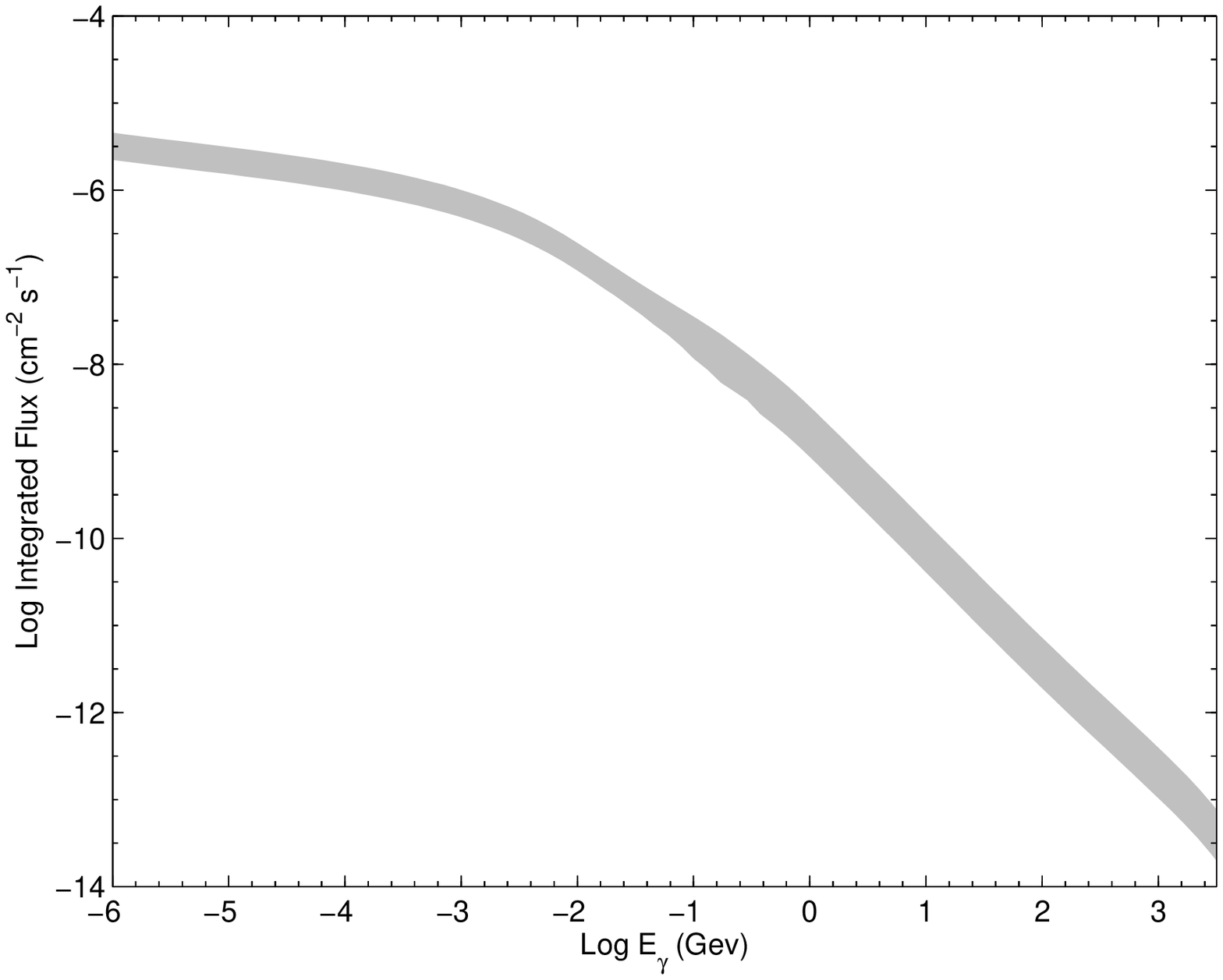,height=8cm,width=8cm,clip=}
\caption{Integrated high energy emission from the disk region of NGC\,253. 
The total integrated emission from the disk region of NGC\,253 is shown 
in the grey region, reflecting uncertainties in the obesrvationallly-deduced 
parameters.}
\end{figure}

The TeV fluxes predicted here are below the upper limit set by the H.E.S.S 
Cherenkov telescope. Treated as a point source, analysis of the H.E.S.S 
measurements (Aharonian \ea 2005) yielded an upper limit $f(\geq 300\,{\rm 
GeV}) \leq 1.9\times 10^{-12}$ cm$^{-2}$ s$^{-1}$ at the 99\% confidence level. 
This limit is higher than the flux predicted here, $f(\geq 300\,{\rm GeV})
\simeq (8.9^{+8.1}_{-4.0})\times 10^{-13}$ cm$^{-2}$ s$^{-1}$.

\section{Discussion}

We have performed a detailed calculation of the steady state 
spectro-spatial distributions of energetic electrons and protons in 
NGC\,253. Our treatment is based on a numerical solution of the 
diffusion-convection equation, following the particles from their 
source region throughout the disk as they lose energy and propagate 
into the interstellar space. A key feature of our treatment is the 
deduction of the electron spectrum and its normalization in the central 
region directly from the measured radio emission in the SB region. 
Adopting the theoretically predicted injection $N_{p}/N_{e}$ (which 
depends on $q$) in the SB source region, and assuming equipartition to 
hold in this region, we iteratively derive the electron and proton 
densities, and the central value of the magnetic field $B_0$, by fitting 
the predicted synchrotorn flux to the observed radio spectrum. Doing so 
fully determines the particle spectra, which are then evolved by solving 
the diffusion-convection equation to predict the spatial distributions 
throughout the disk. Consequently, with the assumed spatial profile of 
the magnetic field set as specified in Section 2, radio emission from 
the full disk is then fully determined. The fact that the predicted 
radio spectrum matches well the observed emission from the disk is an 
important confirmation of the self-consistency of our treatment, and the 
validity of our assumptions.

Our main interest here has been the estimation of VHE 
fluxes from the source region and the entire disk; to do so we 
determine the particle full radiative spectra, from radio to the 
VHE region, and use the measured radio fluxes from both regions to 
normalize the distributions. 

HE $\gamma$-ray emission from the SB region of NGC\,253 was 
estimated by Domingo-Santamaria \& Torres (2005). Their detailed 
treatment focused on a steady state solution to the kinetic equation 
describing the electron and proton spectral distribution functions 
(without explicit spatial dependence). The source term was directly 
related to SN rates in the injection region, and the particles were 
assumed to be accelerated by SN shocks in a small central region. Energy 
losses and diffusion in energy space were followed in the central 1 kpc 
disk region; escape from this region was also included. Estimated 
fluxes - which obviously depend on the spectral parameters and the 
escape time - are generally comparable to our estimates. Our very 
different treatment includes also spatial diffusion and convection, 
and our estimates include the full disk, not just the central 1 kpc 
region. While VHE emission is mostly from the central disk, emission 
from the outer disk is not negligible and provides additional insight 
on the validity of the convection-diffusion model for energetic particle 
propagation in galaxies.

Clearly, our predicted fluxes depend on several parameters, most 
importantly on the proton to electron ratio in the source region, on 
the magnetic field, gas density, and their spatial profiles. The 
electron density was deduced from the measured radio emission in the 
source region; thus, the evolution of the electron spectrum from this 
region to the central disk region is mostly determined by synchrotron 
losses in the high magnetic field. The uncertainty in the estimated 
level of VHE emission stems largely from the steep dependence of the 
electron density on the field. It is unlikely that the field is 
appreciably higher than our deduced value, $B_{0} \sim 200 \, \mu$G. 
A lower mean field value (by a factor $\zeta$), would result in a 
reduced proton density and, consequently, in a lower rate of $\pi^0$ 
decays. However, for a given measured radio flux the electron 
density would have to be correspondingly higher (by a factor 
$\zeta^{(1+q)/2}$), which for a mean electron index of $\sim 2.7$ 
(in the disk) would result in higher bremsstrahlung and Compton 
yields by nearly the same factor ($\sim \zeta^2$). As can be 
seen in Figure 3, at $\sim 100$ GeV the yield from the latter 
processes is only a factor $\sim 2.5$ lower than emission from 
$\pi^0$ decay. Thus, lowering the mean field value in the source 
region by even as much as a factor of two (i.e., much higher than 
the uncertainty implied from the observational radio flux data) 
would not appreciably change our estimated VHE fluxes. The other 
main uncertainty stems from the linear dependence of the p-p 
interactions (and $\pi^0$ decay yield) on the ambient proton density, 
for which we used the observationally deduced mass in the central disk 
region. 

The range of predicted fluxes, which reflect observational uncertainties 
in these quantities, provide a basis for reliable estimates of the 
feasibility of detecting VHE emission from NGC\,253 with current and near 
future telescope arrays, and detection of HE $\gamma$-ray 
emission with GLAST. Based on the above estimate, the predicted VHE flux 
of NGC\,253 falls below the detection limit of current imaging air 
Cherenkov telescopes (IACTs). For example, H.E.S.S.  -- which is located 
in the Southern hemisphere and can therefore observe NGC\,253 -- has a 
5$\sigma$ sensitivity of $\sim$1.8$\times$10$^{-11}$ 
cm$^{-2}$s$^{-1}$ for the detection of emission above 100 GeV in 50 
hours of observation (e.g., Hinton 2004). Even accounting for the fact 
that this sensitivity limit refers to a Crab-like spectrum [which is 
steeper ($\Gamma \simeq 2.6$) than our predicted NGC\,253 spectrum], 
and for the inherent uncertainty in our flux estimate, we conclude that 
detection of VHE $\gamma$-ray emission from NGC\,253 may be feasible 
with H.E.S.S. only if observed for several hundred hours. [Prospects of 
detection with CANGAROO\,III are even less optimistic due to its lower 
sensitivity by a factor 3--5 (Enomoto et al. 2008).] The likelihood of 
detection will appreciably improve with the upcoming H.E.S.S.-II 
telescope, whose sensitivity at $\geq$100 GeV is expected to be 
$\mincir$0.6$\times$10$^{-11}$ cm$^{-2}$s$^{-1}$, i.e. $\sim$3 
times better than H.E.S.S., and with the next-generation facility (the 
Cherenkov Telescope Array, CTA) whose sensitivity at $\geq$100 GeV is 
expected to be $\mincir$10$^{-12}$ cm$^{-2}$s$^{-1}$, i.e. at least  
$\magcir$20 times better than H.E.S.S. 

It is also of interest to compare the predicted $\geq$100\,MeV flux of 
NGC\,253 with the sensitivity of {\it Fermi}'s Large Area Telescope 
(LAT). Our calculated VHE emission yields an integrated flux 
$f(\geq$100\, MeV)$\simeq$2$\times$10$^{-8}$ cm$^{-2}$s$^{-1}$ for a 
differential spectral photon index of $\sim$2.3. This value matches the 
5$\sigma$ sensitivity of the {\it LAT} for a $\sim$1-month scanning-mode 
operation. Thus, {\it Fermi/LAT} should be largely able to detect NGC\,253 
during its first year of operation.

Detection of VHE $\gamma$-ray emission associated with ongoing star 
formation in NGC\,253 will add significant insight on the enhanced 
energetic electron and proton contents in SBGs, and on their propgation 
in disks of spiral galaxies.
\bs

\section{Acknowledgment}

We thank the referee for a critical review of the originally submitted 
version of this paper.
\bs

\def\ref{\par\noindent\hangindent 20pt}

\noindent
{\bf References}
\vglue 0.2truecm

\ref{\small Aharonian, F.A., et al. 2005, A\&A, 442, 177}
\ref{\small Atwood, W.B., et al. 2009, ApJ, 697, 1071} 
\ref{\small Beck, R., \& Krause, M. 2005, Astron. Nachr., 326, 414}
\ref{\small Bell, A.R. 1978, MNRAS, 182, 443 }
\ref{\small Boomsma et al. 2005, ASPC 331, 247}  
\ref{\small Bradford et al. 2003, \apj 586, 891}
\ref{\small Canzian et al 1988, \apj 333, 157}
\ref{\small Carilli C.L. 1996, \aa 305, 402}
\ref{\small De Angelis, A., Mansutti, O., \& Persic, M. 2008, 
Nuovo Cimento, 31, no.4, 187}
\ref{\small de Cea del Pozo, E., Torres, D., \& Rodriguez Marrero, A. 
2009, arXiv:0901.2688}
\ref{\small Domingo-Santamar\'\i a, E., \& Torres, D.F. 2005, A\&A, 
444, 403}
\ref{\small Enomoto, R., et al. 2008, ApJ, 683, 383}
\ref{\small Goldshmidt, O., \& Rephaeli, Y. 1995, ApJ, 444, 113 }
\ref{\small Heesen, V., Beck, R., Krause, M., Dettmar, R.J. 2008, A\&A, 
494, 563} 
\ref{\small Hinton, J. 2004, New Astron. Rev., 48, 331
\ref{\small Klein U. et al. 1983, \aa 127, 177}
\ref{\small Lerche, I., \& Schlickeiser, R. 1982, MNRAS, 201, 1041 }
\ref{\small Melo et al. 2002, \apj 574, 709}
\ref{\small Moskalenko I.V, \& Strong A.W. 1998, ApJ, 493, 694 }
\ref{\small Moskalenko I.V, Jones, F.C., Mashnik, S.G., Ptuskin, V.S., 
\& Strong, A.W., 2003, ICRC, 4, 1925} 
\ref{\small Persic, M., Rephaeli, Y., \& Arieli, Y. 2008, A\&A, 486, 
143 (PRA)}
\ref{\small Protheroe, R.J., \& Clay, R.W. 2004, PASA, 21, 1 }
\ref{\small Rephaeli, Y. 1988, Comm. Ap., 12, 265 }
\ref{\small Romero, G.E., \& Torres, D.F. 2003, ApJ, 586, 33}
\ref{\small Schlickeiser, R. 2002, Cosmic Ray Astrophysics (Berlin: 
Springer), p.472}
\ref{\small Sreekumar et al. 1994, \apj 426, 105}
\ref{\small Strickland et al. 2002, \apj 568, 689}
\ref{\small Strickland, D.K., Ponman, T.J., \& Stevens, I.R. 1997, A\&A, 
320, 378 }
\ref{\small Timothy et al. 1996, \apj 460, 295}

\ref{\small Ulvestad 2000, \aj 120, 278}
\ref{\small Zirakashvili, V.N., Breitschwerdt, D., Ptuskin, V.S., \& 
Voelk, H.J. 1996, A\&A, 311, 113}

\end{document}